\documentclass[prd,twocolumn,superscriptaddress,showpacs,amsmath,amssymb,longbibliography]{revtex4-1}
\usepackage{xcolor}
\usepackage{graphicx}
\usepackage{hyperref}
\usepackage{bm}

\begin{document}

\title{Spacetime geometry from Dirac spinor theory}

\author{Yuri N. Obukhov}
\email{obukhov@ibrae.ac.ru}
\affiliation{Theoretical Physics Laboratory, Nuclear Safety Institute,
Russian Academy of Sciences, B.Tulskaya 52, 115191 Moscow, Russia}

\author{G.E.~Volovik}
\email{grigori.volovik@aalto.fi}
\affiliation{Low Temperature Laboratory, Aalto University,  P.O. Box 15100, FI-00076 Aalto, Finland}
\affiliation{Landau Institute for Theoretical Physics, acad. Semyonov av., 1a, 142432,
Chernogolovka, Russia}

\date{\today}

\begin{abstract}
The quintet of Dirac $4\times 4$ matrices suggests that the fundamental dimension of the internal (spin) space is $n=5$, instead of the conventional dimension $n=4$. Then instead of the conventional $4\times 4$ tetrads (vierbein), gravity is described in terms of the 5-bein (f\"unfbein or five legs).  We discuss the properties of the spacetime geometry induced from this 5-leg Dirac spinor theory, where the spin connection contains $10\times 4=40$ elements instead of 24 elements in the tetrad formulation of the general relativity theory.
\end{abstract}

\maketitle 

 
\section{Introduction}

In the modern formulation of general relativity the spacetime metric is not the primary object. The reason is that the metric does not describe the interaction between gravity and Dirac fermions. Gravity enters the Dirac equation in terms of the vierbein (or tetrads), the $4\times 4$ matrix $e^\mu_a$, which is the primary object.  The metric is the secondary object -- the bi-linear combination of tetrads. The index $\mu$ refers to coordinates of the curved spacetime, and the index $a$ refers to the internal space, which is the $SO(1,3)$ spin space. 

In more general theories, which combine general relativity with Standard Model, the higher dimensional internal spaces can be considered. The internal space may include the groups of Standard Model and the fermionic families. Examples are the  $SO(1,7)$ and $SO(3,11)$ groups \cite{Nesti2010,MaiezzaNesti2022}, the Clifford algebra $Cl(0,6)$ \cite{WeiLu2024}, higher spin fields \cite{Vasiliev1990,Vasiliev2003,Bekaert2005,Boulanger2023}, etc. The situation when the dimension of the internal space $n$ is larger or smaller than the dimension $D$ of the spacetime takes place also in condensed matter, where the rectangular vielbein naturally emerge \cite{Volovik2022,Volovik2020,Volovik2023}. Such examples suggest the possibility to describe gravity using the rectangular $D\times n$ vielbein with $n \neq D$. 

In principle,  vielbein's legs are not necessary associated with the local axes of spacetime coordinate system. They can be the emergent fields, which have nothing to do with the geometry of spacetime. Examples are the Akama-Diakonov-Wetterich theory \cite{Akama1978,Wetterich2004,Diakonov2011,VladimirovDiakonov2012,ObukhovHehl2012} and superfluid $^3$He in the B-phase \cite{Volovik1990}, where the dynamical vielbein emerge as the bi-linear combinations of the fermionic fields. In this case it is also natural that the dimension of spin space can be larger than the dimension of the coordinate space.

In the Dirac equation in $D=4$, the dimension of the Dirac spinor $\psi$ is 4. If we consider only single fermionic species and ignore the gauge degrees of freedom,  then from the properties of the Dirac matrices it follows that the internal spin space may have dimension  $n=5$. Moreover,  $n=5$ is the largest possible dimension  of the internal spin space for Dirac fermions. This suggests that the spin group $SO(1,4)$ and the corresponding $4\times 5$ rectangular vielbein can be the natural elements of the theory describing the interaction of Dirac fermions with gravity. 

In this paper we consider the 5-bein (f\"unfbein) Dirac spinor theory. Our notations and conventions are as follows. Spacetime coordinates are labeled by indices from the Greek alphabet $\mu, \nu, \dots = 0,1,2,3$, while the spin indices $a, b, \dots = 0,1,2,3,4$ take the values from the Latin alphabet.

\section{F\"unfbein Dirac equation}

The fundamental quintet of the Dirac $4\times 4$ matrices $\Gamma^a$ with $a = 0,1,2,3,4$, which obey the anti-commutation relations,
\begin{equation}\label{GammaD}
\{\Gamma^a , \Gamma^b\}=2\eta^{ab}\,,\quad \eta^{ab}={\rm diag}(1,-1,-1,-1,-1) \,,
\end{equation}
can be introduced in terms of the conventional $4\times 4$ Dirac $\gamma$-matrices. Using Peskin-Schr\"oder notations \cite{Peskin2018} for $\gamma$ matrices in terms of the Pauli matrices $\bm{\tau}$ and $\bm{\sigma}$ one has
\begin{eqnarray}
\Gamma^0 &:=& \gamma^0=\tau_1 \qquad (a=0)\,,\label{5matrices1} \\
\Gamma^a &:=& \gamma^a=i\tau_2 \sigma^a \,\,\,(a=1,2,3)\,, \label{5matrices12} \\
\Gamma^4 &:=& -i\gamma_5=i\tau_3 \,\,\,(a=4) \,,\label{5matrices3} \\
\frac{1}{5!} \varepsilon_{abcde}\Gamma^a\Gamma^b\Gamma^c \Gamma^d \Gamma^e
&=& \Gamma^0\Gamma^1\Gamma^2 \Gamma^3 \Gamma^4 =1\,.\label{5matrices4}
\end{eqnarray}
Here $\varepsilon_{abcde}$ is the five-dimensional totally antisymmetric Levi-Civita tensor; its only nontrivial component is $\varepsilon_{01234} = +1$, hence also $\varepsilon^{01234} = +1$.

In order to formulate the dynamics of a spinor field in the curved four-dimensional spacetime, we combine the $4\times 5$ vielbein $e^\mu_a$ and five matrices $\Gamma^a$, which form the 5-vector in spin space, into a $4\times 4$ Dirac matrices $\gamma^\mu(x) = e^\mu_a(x) \Gamma^a$ which form the 4-vector on the spacetime. The resulting Dirac equation for a fermion with the rest mass $M$ in the curved manifold then reads
\begin{equation}\label{DiracCurved}
\left(i e^\mu_a \Gamma^a \nabla_\mu - M\right)\psi=0\,.
\end{equation}
This wave equation is covariant under arbitrary general coordinate transformations, $x\longrightarrow x(x')$, on the spacetime manifold and under the local $SO(1,4)$ transformations in spin space:
\begin{eqnarray}
e^\mu_a \longrightarrow e^\mu_b\Lambda^b{}_a,&\quad& \psi \longrightarrow U^{-1}\psi,\label{so1}\\ 
U^{-1}\Gamma^a U &=& \Lambda^a{}_b\Gamma^b.\label{so2}
\end{eqnarray}
The orthogonal $5\times 5$ matrices $\Lambda^a{}_b(x)$ (such that $\Lambda^a{}_c\Lambda^b{}_d\eta^{cd} = \eta^{ab}$) generate the $SO(1,4)$ spin transformations $U(x)$ by means of the operators
\begin{equation}
S^{ab}=\frac{i}{4}[\Gamma^a,\Gamma^b]\,.
\label{algebra}
\end{equation}
The latter determine the covariant spinor derivatives in terms of the spin connection $\omega_{ab\mu} = -\,\omega_{ba\mu}$. Explicitly, for the fermion wave function one has:
\begin{equation}
\nabla_\mu \psi =\partial_\mu\psi + \omega_\mu\psi,\qquad \omega_\mu = \frac{i}{2}\omega_{ab\mu} S^{ab}.
\label{GradientPsi}
\end{equation}
Note that the $SO(1,4)$ spin connection contains $10\times 4=40$ elements instead of 24 elements in the tetrad formulation of the general relativity (GR) theory. By evaluating the commutator of covariant derivatives,
\begin{equation}\label{ddpsi}
\left(\nabla_\mu\nabla_\nu - \nabla_\nu\nabla_\mu\right)\psi = \frac{i}{2}\Omega_{ab\mu\nu} S^{ab}\psi,
\end{equation}
we derive the spin field strength, or a {\it spin curvature}:
\begin{equation}\label{Fso}
\Omega^a{}_{b\mu\nu} = \partial_\mu \omega^a{}_{b\nu} - \partial_\nu \omega^a{}_{b\mu} 
+ \omega^a{}_{c\mu}\omega^c{}_{b\nu} - \omega^a{}_{c\nu}\omega^c{}_{b\mu},
\end{equation}

It is worthwhile to recall the key algebraic relations:
\begin{eqnarray}
\left[\Gamma^a, S^{bc}\right] &=& i\left(\eta^{ab}\Gamma^c - \eta^{ac}\Gamma^b\right),\label{comm1}\\
\left\{\Gamma^a, S^{bc}\right\} &=& \varepsilon^{abcde}\,S_{de}.\label{comm2}
\end{eqnarray}
Furthermore, the $SO(1,4)$ generators (\ref{algebra}) satisfy
\begin{eqnarray}
\left[S^{ab}, S^{cd}\right] &=& i\bigl(-\,S^{ac}\eta^{bd} + S^{ad}\eta^{bc} \nonumber\\
&& +\,S^{bc}\eta^{ad} - S^{bd}\eta^{ac}\bigr),\label{comm3}\\
\left\{S^{ab}, S^{cd}\right\} &=& {\frac 12}\left(\eta^{ac}\eta^{bd} - \eta^{ad}\eta^{bc}
+ \varepsilon^{abcde}\,\Gamma_e\right).\label{comm4}
\end{eqnarray}

\section{Spacetime geometry induced by $SO(1,4)$ spin structure}\label{spacetime}

We treat $\left\{e^\mu_a, \omega^a{}_{b\mu}\right\}$ as the fundamental internal spin variables in this approach. Let us demonstrate that they give rise to an {\it induced geometrical structure} on the spacetime manifold, that encompasses the metric $g^{\mu\nu}$ and the linear connection $\Gamma^\alpha{}_{\beta\mu}$.

The traditional $4\times 4$ spacetime metric is expressed in terms of the $4\times 5$ vielbein in a pretty much the same way as in the tetrad GR:
\begin{equation}\label{MetricDirac}
g^{\mu\nu} = e^\mu_a e^\nu_b \eta^{ab}. 
\end{equation}
This metric $g^{\mu\nu}$ then enters the usual equations for the electromagnetic gauge fields and for the other bosonic fields, which do not depend on spin degrees of freedom. Obviously, the spacetime metric (\ref{MetricDirac}) is a symmetric covariant second rank tensor field, and it is invariant under arbitrary local $SO(1,4)$ transformations (\ref{so1}) in the spin space.

Provided the spacetime metric is nondegenerate (i.e., $\det g^{\mu\nu} \neq 0$), we can construct the inverse tensor field $g_{\mu\nu}$. However, since (\ref{MetricDirac}) relates 4- and 5-dimensional spaces, the rectangular $4\times 5$ vielbein $e^\mu_a$ cannot be inverted. Nevertheless, we can introduce a $5\times 4$ matrix
\begin{equation}\label{einv}
e^a_\mu := \eta^{ab} g_{\mu\nu} e^\nu_b,
\end{equation}
which by construction is {\it semi-inverse} of the original vielbein, in the sense that
\begin{equation}\label{ee}
e^a_\mu e^\nu_a = \delta^\nu_\mu.
\end{equation}
However,
\begin{equation}\label{Pi}
e^a_\mu e^\mu_b = \Pi^a{}_b \neq \delta^a_b.
\end{equation}
From the definition, we prove that this is an idempotent object (hence, a projector),
\begin{equation}\label{Pi2}
\Pi^a{}_c \Pi^c{}_b = \Pi^a{}_b.
\end{equation}
In general, it depends on the spacetime coordinates, $\Pi^a{}_b = \Pi^a{}_b(x)$.

Despite its peculiar properties above, the semi-inverse matrix is a quite valuable variable because with its help one can define the new hybrid object:
\begin{equation}\label{theta}
\Theta^a{}_{\mu\nu} = \partial_\mu e_\nu^a - \partial_\nu e_\mu^a
+ \omega^a{}_{b\mu}e_\nu^b - \omega^a{}_{b\nu}e_\mu^b\,.
\end{equation}
Together, the spin curvature (\ref{Fso}) and the {\it spin torsion} (\ref{theta}) play the role of the generalized structure relations in spin space. 

After these preliminaries, we are in a position to discuss the linear connection $\Gamma^\alpha{}_{\beta\mu}$ on the curved spacetime manifold. Let us demonstrate how it arises from the crucial  consistency condition of the internal $SO(1,4)$ spin structure with the geometrical structure of the spacetime:
\begin{equation}\label{dg0}
\nabla_\mu\gamma^\nu = \partial_\mu\gamma^\nu + \Gamma^\nu{}_{\lambda\mu}\gamma^\lambda
+ \omega_\mu\gamma^\nu - \gamma^\nu\omega_\mu = 0.
\end{equation}
Making use of (\ref{GradientPsi}) and (\ref{comm1}), we recast (\ref{dg0}) into an equivalent form
\begin{equation}\label{de0}
\nabla_\mu e^\nu_a = \partial_\mu e^\nu_a + \Gamma^\nu{}_{\lambda\mu} e^\lambda_a - \omega^b{}_{a\mu}e^\nu_b = 0.
\end{equation}
Since the vielbein cannot be inverted, one cannot solve this to find the spin connection $\omega^b{}_{a\mu}$ in terms of the spacetime geometrical structures. However, with the help of the semi-inverse matrix $e^a_\mu$ we straightforwardly derive from (\ref{de0}):
\begin{equation}\label{conom}
\Gamma^\alpha{}_{\beta\mu} = e^\alpha_a\omega^a{}_{b\mu}e_\beta^b + e^\alpha_a \partial_\mu e_\beta^a.
\end{equation}

Summarizing, we have demonstrated that the $SO(1,4)$ spin structure {\it induces} the geometrical structure of the spacetime,
\begin{equation}\label{map}
\left( e^\mu_a, \omega^a{}_{b\mu}\right) \Longrightarrow \left( g^{\mu\nu}, \Gamma^\alpha{}_{\beta\mu} \right),
\end{equation}
by means of (\ref{MetricDirac}) and (\ref{conom}).

\section{Properties of induced spacetime geometry}\label{geom}

In general, the metric-affine manifold, endowed with the metric and the linear connection $\left\{g^{\mu\nu}, \Gamma^\alpha{}_{\beta\mu}\right\}$, is characterized by the curvature, torsion and nonmetricity \cite{MAG}. 

By differentiating (\ref{de0}) we derive the integrability condition of this consistency relation: 
\begin{equation}\label{RF}
R^\alpha{}_{\beta\mu\nu} e^\beta_a - \Omega^b{}_{a\mu\nu} e^\alpha_b = 0, 
\end{equation}
which relates the spacetime curvature
\begin{equation}\label{curv}
R^\alpha{}_{\beta\mu\nu} = \partial_\mu \Gamma^\alpha{}_{\beta\nu} - \partial_\nu \Gamma^\alpha{}_{\beta\mu} 
+ \Gamma^\alpha{}_{\lambda\mu}\Gamma^\lambda{}_{\beta\nu} - \Gamma^\alpha{}_{\lambda\nu}\Gamma^\lambda{}_{\beta\mu},
\end{equation}
and the $SO(1,4)$ spin curvature (\ref{Fso}). Thereby, with the help of the semi-inverse vielbein from (\ref{RF}) we find the spacetime curvature in terms of the spin curvature:
\begin{equation}\label{RF1}
R^\alpha{}_{\beta\mu\nu} =  \Omega^a{}_{b\mu\nu} e^\alpha_a e_\beta^b\,. 
\end{equation}
By contraction, we derive the corresponding relations for the Ricci tensor and the curvature scalar:
\begin{eqnarray}
R_{\mu\nu} &=& R^\lambda{}_{\mu\lambda\nu} =  e^\lambda_a e_\beta^b\,\Omega^a{}_{b\lambda\nu},\label{Ric}\\
R &=& g^{\mu\nu}R_{\mu\nu} = e^\mu_a e^\nu_b\,\Omega^{ab}{}_{\mu\nu}.\label{RS}
\end{eqnarray}

Using (\ref{MetricDirac}) and (\ref{conom}), we straightforwardly verify that the spacetime connection $\Gamma^\alpha{}_{\beta\mu}$ is compatible with the metric:
\begin{equation}\label{nonmet0}
\nabla_\mu g^{\alpha\beta} = \partial_\mu g^{\alpha\beta} +
\Gamma^\alpha{}_{\lambda\mu} g^{\lambda\beta} + \Gamma^\beta{}_{\lambda\mu} g^{\alpha\lambda} = 0.
\end{equation}
In other words, the induced spacetime geometry has the {\it vanishing nonmetricity}.

However, the induced spacetime torsion
\begin{equation}\label{tor}
T^\alpha{}_{\mu\nu} = \Gamma^\alpha{}_{\nu\mu} - \Gamma^\alpha{}_{\mu\nu} = e^\alpha_a\Theta^a{}_{\mu\nu}.
\end{equation}
can be nontrivial, in general, which is explained by the larger number of components ($4\times 10 = 40$) of the $SO(1,4)$ spin connection. 

Multiplying the generalized Dirac equation (\ref{DiracCurved}) by $(-i e^\mu_a \Gamma^a \nabla_\mu - M)$, with an account of the key consistency condition (\ref{dg0}), one obtains the generalized Klein-Gordon equation:
\begin{equation}
(g^{\mu\nu} \nabla_\mu  \nabla_\nu + {\mathcal T}^\lambda\nabla_\lambda + {\mathcal R} + M^2)\psi =0\,.
\label{DiracSq}
\end{equation}
Here the operators
\begin{eqnarray}
{\mathcal T}^\lambda &=& iS^{ab}\,e^\mu_a e^\nu_b e_c^\lambda\,\Theta^c{}_{\mu\nu},\label{TD}\\
{\mathcal R} &=& {\frac 12}\,S^{ab}S^{cd}\,e^\mu_a e^\nu_b\,\Omega_{cd\mu\nu} \label{RSSF}
\end{eqnarray}
are determined by the $SO(1,4)$ spin torsion and curvature. Using (\ref{comm3}) and (\ref{comm4}) we can reduce (\ref{RSSF}) to
\begin{equation}\label{Lich}
{\mathcal R} = {\frac R4} - iS^{bc}\Omega^a{}_{c\mu\nu}\,e^\mu_a e^\nu_b
+ {\frac 12}\,\varepsilon^{ab}{}_{cde}\,\Gamma^e \Omega^{cd}{}_{\mu\nu}\,e^\mu_a e^\nu_b\,,
\end{equation}
which generalizes the earlier results \cite{Lichnerowicz,Duff}. The two last terms are trivial in Einstein's GR.

\section{Understanding rectangular vielbein}
\subsection{Asymmetric splitting of vielbein}

The structure of an arbitrary $4\times 5$ vielbein can be quite complicated, in general. But from the formal mathematical point of view, this is a rectangular matrix
\begin{equation}\label{emat}
e^\mu_a = \left(\begin{array}{cccc|c} e^0_0 & e^0_1 & e^0_2 & e^0_3 & e^0_4 \\
e^1_0 & e^1_1 & e^1_2 & e^1_3 & e^1_4 \\ e^2_0 & e^2_1 & e^2_2 & e^2_3 & e^2_4 \\
e^3_0 & e^3_1 & e^3_2 & e^3_3 & e^3_4 \end{array}\right),
\end{equation}
and it is natural to split this matrix into a square $4\times 4$ block $e^\mu_\alpha$ (with $\alpha = 0,1,2,3$) and treat the last column $k^\mu := e^\mu_4$ as an additional 4-vector element.

Using this decomposition $e^\mu_a = \left\{e^\mu_\alpha, k^\mu\right\}$, we recast the Dirac equation (\ref{DiracCurved}) into an equivalent form
\begin{equation}\label{DiracL}
\left(i e^\mu_\alpha \gamma^\alpha \nabla_\mu + \gamma_5k^\mu\nabla_\mu - M\right)\psi = 0\,.
\end{equation} 
We immediately recognize this as a particular case of the so called standard-model extension (SME), where the variable $k^\mu$ manifests the Lorentz-violating effects, see Kostelecky et al \cite{Alan:1997,Alan:1998}. This appears to be a natural result, since the spin group $SO(1,4)$ is indeed an extension of the Lorentz $SO(1,3)$ symmetry. 

On the other hand, inserting  $e^\mu_a = \left\{e^\mu_\alpha, k^\mu\right\}$ into (\ref{MetricDirac}), we find for the induced spacetime metric  
\begin{equation}\label{KS}
g^{\mu\nu} = {\stackrel {(0)} g}{}^{\mu\nu} - k^\mu k^\nu,
\end{equation}
where the first term, ${\stackrel {(0)} g}{}^{\mu\nu} = e^\mu_\alpha e^\nu_\beta \eta^{\alpha\beta}$, is constructed with the standard four-dimensional flat Minkowski $\eta^{\alpha\beta} = {\rm diag}(1,-1,-1,-1)$. The representation (\ref{KS}) is known as the generalized Kerr-Schild ansatz, which is a powerful tool widely used in gravity theory to generate exact solutions of the gravitational field equations, see \cite{exact,Gurses,ppwave}. Of special interest is the case when $k^\mu$ is null (lightlike) vector field with respect to both metrics $g^{\mu\nu}k_\mu k_\nu = {\stackrel {(0)} g}{}^{\mu\nu}k_\mu k_\nu = 0$. The representation (\ref{KS}) may be particularly useful for the study of physical effects in the Kerr geometry (with a nontrivial torsion, in general), such as the superradiance \cite{Unruh,Guven}, for example, however, this goes beyond the scope of the present paper.

\subsection{Symmetric five-leg vielbein}

The Euclidean four-dimensional space with $g^{\mu\nu} = \delta^{\mu\nu} = \rm{diag}(1,1,1,1)$ (where $\mu, \nu = 0,1,2,3$), $\Gamma^a = i\gamma^a$ (for $a = 1,2,3$) and $\Gamma^4 = \gamma_5$, can be constructed using the symmetric five-leg vielbein -- the vertices of the 4-simplex:
\begin{equation}\label{fiveleg}
e^\mu_a = \frac{1}{2}\left(\begin{array}{ccccc} 
1 & 1 & -1 & -1& 0\\ 1 & -1 & 1 & -1& 0 \\ 1 & -1 & -1 & 1 & 0\\
-\frac{1}{\sqrt{5}}& -\frac{1}{\sqrt{5}} &-\frac{1}{\sqrt{5}}&
-\frac{1}{\sqrt{5}} & \frac{4}{\sqrt{5}} \end{array}\right).
\end{equation}
Note that the internal spin group is the Euclidean $SO(5)$. The hyper-pyramid is fully symmetric:
\begin{eqnarray}
e^\mu_0 + e^\mu_1 + e^\mu_2 + e^\mu_3 + e^\mu_4 &=& 0,\quad \mu = 0,1,2,3,\label{hyperpyramid1}\\
\delta_{\mu\nu}\,e^\mu_a e^\nu_b &=& \frac{4}{5},\qquad a= b\,,\label{hyperpyramid2}\\
\delta_{\mu\nu}\,e^\mu_a e^\nu_b &=& -\,{\frac{1}{5}},\quad a\neq b\,.\label{hyperpyramid3}
\end{eqnarray}
This has natural geometrical connection to two golden ratios, $\phi_+$ and $\phi_-$:
\begin{eqnarray}
\phi_\pm = \frac{1\pm \sqrt{5} }{2}\,,\label{golden1}\\
\phi_+ + \phi_-=1 \,,\quad \phi_+ - \phi_-=  \sqrt{5} \,.\label{golden2}
\end{eqnarray}

\subsection{Vacuum quasicrystal}

The rectangular vielbein emerges also in such condensed matter systems as quasicrystal -- the structure in the $(D=3)$-dimensional space, which is obtained from a regular space crystal in dimension $n > D$  by dimensional reduction \cite{NissinenVolovik2018}. The quasicrystal structure can be described as a system of $n$ crystallographic planes, $X^a(x)=2\pi N^a$, $N^a \in \mathbb{Z}$ with $a=1,2,\dots,n$. The elasticity properties of the quasicrystals are described by the so-called elasticity vielbein, which are the gradients of the phase functions:
\begin{equation}\label{reciprocal}
e^{~a}_i(x)= \partial_i X^a(x),\quad i=1,2,3, \quad a=1,2,\dots,n\,.
\end{equation}

The five-leg vielbein in Eq.(\ref{fiveleg}) can be considered in terms of the quasicrystalline vacuum, which is obtained by reduction from the regular 5D crystal to the 4D space:
\begin{equation}\label{reciprocalQuasi}
e^{~a}_\mu= \partial_\mu X^a(x),\quad \mu=0,1,2,3, \quad a=0,1,2,3,4 \,.
\end{equation}
In the flat spacetime, this gravitational quasicrystalline vacuum looks as uniform, since $g^{\mu\nu} = \delta^{\mu\nu} = \rm{diag}(1,1,1,1)$.

In the elasticity vielbein approach, the curvature and torsion are obtained by introduction of the topological defects into the crystal structure -- dislocations and disclinations \cite{DzyaloshinskiiVolovik1980,Bilby1956,Kroner1960}.

\section{de Sitter spacetime and de Sitter group $SO(1,4)$}

The $4\times 5$ vielbein can be also obtained if one considers the de Sitter spacetime as the $(1+3)$-hypersurface in the $(1+4)$-dimensional Minkowski space-time $M_{1,4}$. With the Cartesian coordinates ${\cal X}^a, a = 0,1,2,3,4,$ and the line element $ds^2 = \eta_{ab}d{\cal X}^ad{\cal X}^b$ on $M_{1,4}$, the de Sitter space $\Sigma_{1,3}$ can be embedded in it as the hyperboloid \cite{DS}
\begin{equation}\label{hy}
\eta_{ab}{\cal X}^a{\cal X}^b = - \,\ell^2.
\end{equation}
The constant parameter $\ell$ is called a {\it radius} of the de Sitter space $\Sigma_{1,3}$. The embedding (\ref{hy}) can be, for example, conveniently described in parametric form by the Cartesian coordinates $x^\mu = (t, x, y, z)$: 
\begin{eqnarray}\label{X0}
{\cal X}^0 &=& \ell\,f(r)\,\mathbb{S}(t),\ {\cal X}^4 = \ell\,f(r)\,\mathbb{C}(t),\\
{\cal X}^1 &=& x,\quad {\cal X}^2 = y,\quad {\cal X}^3 = z,\label{X1}
\end{eqnarray}
where $r^2 = x^2 + y^2 + z^2$ and the functions
\begin{equation}\label{fCS}
f = \sqrt{1 - {\frac {r^2}{\ell^2}}},\quad \mathbb{C} = \cosh(t/\ell),
\quad \mathbb{S} = \sinh(t/\ell).
\end{equation}
Thereby, on the hypersurface (\ref{hy}) the metric is induced
\begin{equation}\label{dsds}
ds^2 = g_{\mu\nu}dx^\mu dx^\nu,\qquad g_{\mu\nu} = e_\mu^a e_\nu^b \eta_{ab},
\end{equation}
by the $5\times 4$ vielbein of the embedding (\ref{X0}), (\ref{X1}):
\begin{equation}\label{eDS}
e_\mu^a = {\frac {\partial {\cal X}^a}{\partial x^\mu}} = \left(\begin{array}{cccc}f\mathbb{C} &
-{\frac {x\,\mathbb{S}}{\ell\,f}} & -{\frac {y\,\mathbb{S}}{\ell\,f}} & -{\frac {z\,\mathbb{S}}{\ell\,f}}\\
0 & 1 & 0 & 0 \\ 0 & 0 & 1 & 0 \\ 0 & 0 & 0 & 1 \\ f\mathbb{S} &
-{\frac {x\,\mathbb{C}}{\ell\,f}} & -{\frac {y\,\mathbb{C}}{\ell\,f}} & -{\frac {z\,\mathbb{C}}{\ell\,f}}
\end{array}\right).
\end{equation}
The resulting induced metric (\ref{dsds},
\begin{equation}\label{metDS}
g_{\mu\nu} = \left(\begin{array}{cc} f^2 & 0 \\ 
0 & -\delta_{ij} - {\frac {x_ix_j}{\ell^2f^2}}\end{array}\right),\quad i,j = 1,2,3,
\end{equation}
describes $\Sigma_{1,3}$ as a homogeneous four-dimensional spacetime of constant curvature $R^\alpha{}_{\beta\mu\nu} = {\frac {1}{\ell^2}}\left(\delta^\alpha_\nu g_{\beta\mu} - \delta^\alpha_\mu g_{\beta\nu}\right)$. Transforming from the Cartesian $(x,y,z)$ coordinates to the spherical ones $(r,\theta,\phi)$, we can recast (\ref{dsds})-(\ref{metDS}) into a familiar de Sitter line element
\begin{equation}\label{ds2}
ds^2 = \left(1 - {\frac {r^2}{\ell^2}}\right) dt^2 - {\frac {dr^2}{1 - {\frac {r^2}{\ell^2}}}}
- r^2(d\theta^2 + \sin^2\theta d\phi^2).
\end{equation}

Geometrically, the orthogonal group $SO(1,4)$ is the group of motions of the de Sitter space, and the latter arises as homogeneous $\Sigma_{1,3} = SO(1,4)/SO(1,3)$ space. Hence, it is common to call $SO(1,4)$ a {\it de Sitter group}. Choosing the point ${\cal X}^a_0 = \left\{0,0,0,0,\ell\right\} \in \Sigma_{1,3}$ as a center, we have as its stabilizer the Lorentz group $SO(1,3)$ formed by $5\times 5$ matrices ${\stackrel L\Lambda}{}^a{}_b = \left(\begin{array}{cc} L & 0 \\ 0 & 1\end{array}\right)$, with $L \in SO(1,3)$. Accordingly, the action of the 10-parameter de Sitter group $SO(1,4)$ is naturally parametrized by the product of $5 \times 5$ matrices
\begin{equation}\label{LLS}
\Lambda = {\stackrel L\Lambda}\,{\stackrel S\Lambda}, 
\end{equation}
where ${\stackrel L\Lambda}$ is an arbitrary 6-parameter Lorentz rotation, and the matrix ${\stackrel S\Lambda}$ describes a 4-parameter {\it de Sitter boost} that maps the center ${\cal X}^a_0 \in \Sigma_{1,3}$ to an arbitrary point ${\cal X}^a = \ell t^a \in \Sigma_{1,3}$, with $\eta_{ab}t^at^b = -\,1$. Explicitly,
\begin{equation}\label{boost}
{\stackrel S\Lambda}{}^a{}_b = \left(\begin{array}{cc} S^\alpha{}_\beta & t^\alpha \\ t_\beta & t^4
\end{array}\right),\quad S^\alpha{}_\beta = \delta^\alpha_\beta + {\frac {t^\alpha t_\beta}{1 + t^4}}, 
\end{equation}
where $\alpha, \beta = 0,1,2,3$, and $t_\alpha = \eta_{\alpha\beta}t^\beta$, with the 4-dimen\-sional Minkowski metric $\eta_{\alpha\beta} = {\rm diag}(1,-1,-1,-1)$.

In the limit of $\ell\longrightarrow\infty$, the de Sitter space $\Sigma_{1,3}$ reduces to the flat spacetime $R^\alpha{}_{\beta\mu\nu} = 0$ with (\ref{metDS}) becoming the standard Minkowski metric, whereas the de Sitter group $SO(1,4)$ reduces to the corresponding group of motions of the Minkowski spacetime $M_{1,3}$, i.e, the 10-parameter Poincar\'e group. In terms of the Lie algebras, the reduction of $so(1,4)$ to the Poincar\'e algebra is known as the In\"on\"u-Wigner contraction \cite{Inonu}.

Interestingly, Dirac \cite{Dirac} proposed an alternative wave equation for a spin ${\frac 12}$ particle in the de Sitter spacetime, which was later revisited by G\"ursey and Lee \cite{Gursey}, by making use of the 5D Clifford algebra (\ref{GammaD}) and the embedding geometry (\ref{hy}). Despite certain formal similarities, that spinor wave equation is quite different from our approach based on the $4\times 5$ vielbein.

\section{Conclusion}

The generalized Dirac equation (\ref{DiracCurved}) suggests a nontrivial extension of the Lorentz $SO(1,3)$ spin group to the de Sitter group $SO(1,4)$, introducing the rectangular $4\times 5$ vielbein $e^\mu_a$ as a new fundamental variable along with the spin connection $\omega_{ab\mu} = -\,\omega_{ba\mu}$ that is necessary to provide the covariance of the Dirac wave equation under arbitrary general coordinate transformations on the spacetime manifold and under the local de Sitter $SO(1,4)$ transformations (\ref{so1}) in spin space.

Here we demonstrated that the fundamental spin variables $\left\{e^\mu_a, \omega^a{}_{b\mu}\right\}$ give rise to an induced geometrical structure on the spacetime manifold (\ref{map}), and determine the metric $g^{\mu\nu}$ and the linear connection $\Gamma^\alpha{}_{\beta\mu}$ via (\ref{MetricDirac}) and (\ref{conom}), respectively. The resulting Riemann-Cartan spacetime geometry is characterized by the vanishing nonmetricity (\ref{nonmet0}), whereas the nontrivial spacetime curvature and torsion are constructed in terms of the spin curvature and the spin torsion via (\ref{RF1}) and (\ref{tor}).

The use of rectangular vielbeins introduces a new field-theoretic approach for the discussion of the physically important questions such as the chirality issue and the possible Lorentz symmetry violation, that allows for an interesting mixing of internal and spacetime symmetries, thereby naturally taking gravity into account, however, without bringing in extra spacetime dimensions of the Kaluza-Klein type. Among other prospective applicaitons of rectangialr vielbeins, it is also worthwhile to mention the study of their potential role in condensed matter physics \cite{Volovik1990,Volovik2022,Volovik2020,Volovik2023}.

\section*{Acknowledgments}

GEV thanks Sergey Bondarenko, Thibault Damour, Michael Stone, Sergey Vergeles, Wei Lu, Andrei Zelnikov and Mikhail Zubkov for discussion.

\end{document}